\documentclass[multphys,vecphys]{svmult}

\usepackage{makeidx}         % allows index generation
\usepackage{graphicx}        % standard LaTeX graphics tool
\usepackage{multicol}        % used for the two-column index
\usepackage[bottom]{footmisc}% places footnotes at page bottom
\makeindex             % used for the subject index
\begin{document}

\title*{Simulating Diffuse Light in Galaxy Clusters}
% Use \titlerunning{Short Title} for an abbreviated version of
% your contribution title if the original one is too long
\author{Craig S. Rudick\inst{1} \and J. Christopher Mihos\inst{1} \and
  Cameron McBride\inst{1}\ \inst{2}}
% Use \authorrunning{Short Title} for an abbreviated version of
% your contribution title if the original one is too long
\institute{Department of Astronomy, Case Western Reserve University,
10900 Euclid Ave, Cleveland, OH 44106
\texttt{craig@fafnir.astr.cwru.edu, mihos@case.edu}
\and Now in the Department of Physics and Astronomy, University of Pittsburgh}
%
% Use the package "url.sty" to avoid
% problems with special characters
% used in your e-mail or web address
%
\maketitle

\begin{abstract}
Using $N$-body simulations, we have modeled the production and evolution of
low surface brightness, diffuse intra-cluster light (ICL) in galaxy clusters.
By creating simulated observations of the clusters we have measured
the evolution of the ICL luminosity throughout
the dynamical history of the clusters.  We find that ICL production
tends to occur in short, discrete events, which correlate very
strongly with strong, small-scale interactions and accretions between
groups within the clusters.
\end{abstract}

\section{Simulated Deep Imaging of Galaxy Clusters}
\label{rud:sec:simimages}
Diffuse, intra-cluster starlight (ICL) has been observed in numerous
galaxy clusters in the local universe (see \cite{rud:mihos} for more on ICL
observations).  As the product of tidal interactions between cluster
galaxies, this very low surface brightness material has the potential
to reveal a great deal about the nature of dynamical interactions
within the cluster. In order to better understand the processes which
create the observed ICL, we have created $N$-body simulations of the
evolution of luminous galaxies in clusters, in the context of a
$\Lambda CDM$ universe (see \cite{rud:rudick2006} for more information on
our simulation techniques).  We have modeled
three such clusters, each approximately $10^{14} M_{\odot}$ (referred to
as clusters C1, C2, and C3 respectively, henceforth).

In order to examine the evolution of the ICL in our clusters, we have
created simulated deep photometric images of the clusters evolving
from $z=2$ to $z=0$.
First, we smoothed the distribution of discrete particles
into a continuous mass distribution.  This was done by first projecting
the three-dimensional particle positions onto a two-dimensional
plane, then smoothing the particles using an adaptive two-dimensional
Gaussian smoothing kernel.

We obtained luminosity distributions of the clusters by applying a
mass-to-light ratio ($M/L$) to the smoothed mass distribution.
Our $N$-body simulations, however, have not
modeled gas physics, and therefore star formation and stellar evolution, in
any way.  We have therefore chosen to adopt a global $M/L$ of $5
M_{\odot}/L_{\odot}$ in the $V$-band, a value characteristic of older stellar
populations in the local universe, such as we expect to dominate in
the dense cluster environments we have simulated.  This $M/L$ has been
applied to all images at all evolutionary times.  While it is
certainly unphysical to expect to see similar stellar populations at
$z=2$ and $z=0$, this method has several advantages which aid in the
interpreting dynamical changes within the cluster.  Most
importantly, the global $M/L$ means that all changes seen in the
luminosity distribution as the cluster evolves are the direct result
of gravitational dynamics alone.  Secondly, we are not attempting to
simulate cosmological observations of the clusters at distant
redshifts, but rather observing the clusters in their given dynamical
state as they would appear in the local universe.

\section{Evolution of the ICL}

\subsection{Defining ICL}
Most previous theoretical work has defined ICL as stellar particles
which are bound to the cluster potential, but unbound to any
individual galaxy in the cluster
(e.g. \cite{rud:willman2004},\cite{rud:sommerlarsen2005}).  In general,
however, the 
specific binding energy of individual stars is not a readily
observable feature of the ICL.  We have therefore adopted a more
observationally tractable quantity as our definition of ICL: luminosity which 
is at surface brightness fainter than 26.5 mag/sq.arcsec.  We chose
$\mu_V=26.5$ mag/sq.arcsec specifically as our surface brightness
cutoff because in both our simulated images and in direct observations
(\cite{rud:feldmeier2002}, \cite{rud:mihos2005}) this appeared to be
approximately the surface 
brightness limit where the isophotal contours no longer simply outlined
those of higher galaxian surface brightness, but seemed to take on
a distinct morphology.

\subsection{Changes in ICL Luminosity}
\begin{figure}
\centering
\includegraphics[width=\textwidth]{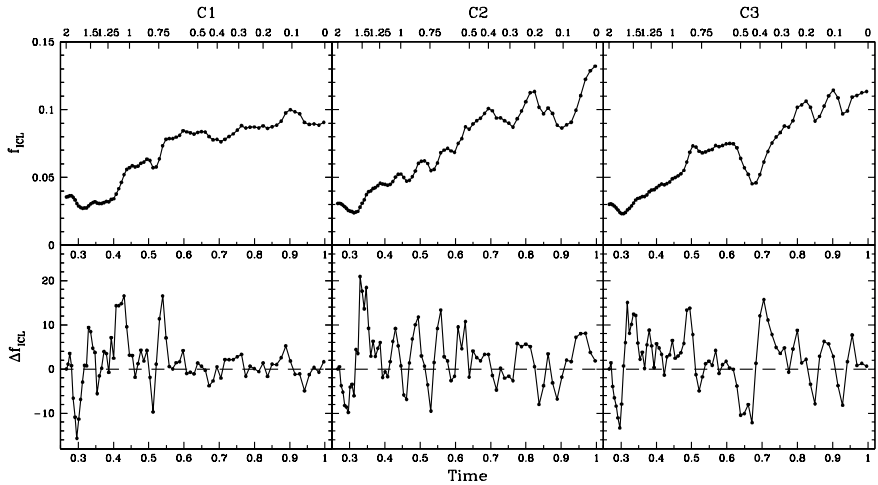}
\caption{\emph{Top: } The fraction of luminosity in each cluster
  which is at ICL surface brightness ($\mu_V>26.5$ mag/sq.arcsec) as a
  function of time ($f_{ICL}$) for each of the three clusters.  The
  top axis shows the corresponding nominal redshift.
  \emph{Bottom:} The fractional change in ICL luminosity per unit time
  as a function of time ($\Delta f_{ICL}$).
  \label{rud:fig:ficl}}
\end{figure}

Using our observational definition of ICL, we have calculated the
fraction of each cluster's luminosity which is at ICL surface
brightness ($f_{ICL}$) as a function of time \footnote{The units of
  time used are such that $t=1$ is the current age of the universe.},
as shown on the top row of Fig. \ref{rud:fig:ficl}.  The most
immediately obvious feature of this analysis is that, in general, the
fraction of ICL luminosity increases as a function of time  in each
of the clusters.  However, it is apparent that this increasing ICL
fraction is far from smooth or  monotonic, including several examples
of large increases in $f_{ICL}$ on short timescales, as well as significant,
extended decreases. 
In order to better understand the changing ICL luminosity fraction, we have
calculated the fractional change in $f_{ICL}$ per unit time ($\Delta
f_{ICL}$), plotted as a function of time on the bottom row of
Fig. \ref{rud:fig:ficl}.  This plot shows even more clearly that
increases in the ICL luminosity fraction tend to come in short,
discrete events, often preceded by significant $f_{ICL}$ decreases.

\begin{figure}
\centering
\includegraphics[width=\textwidth]{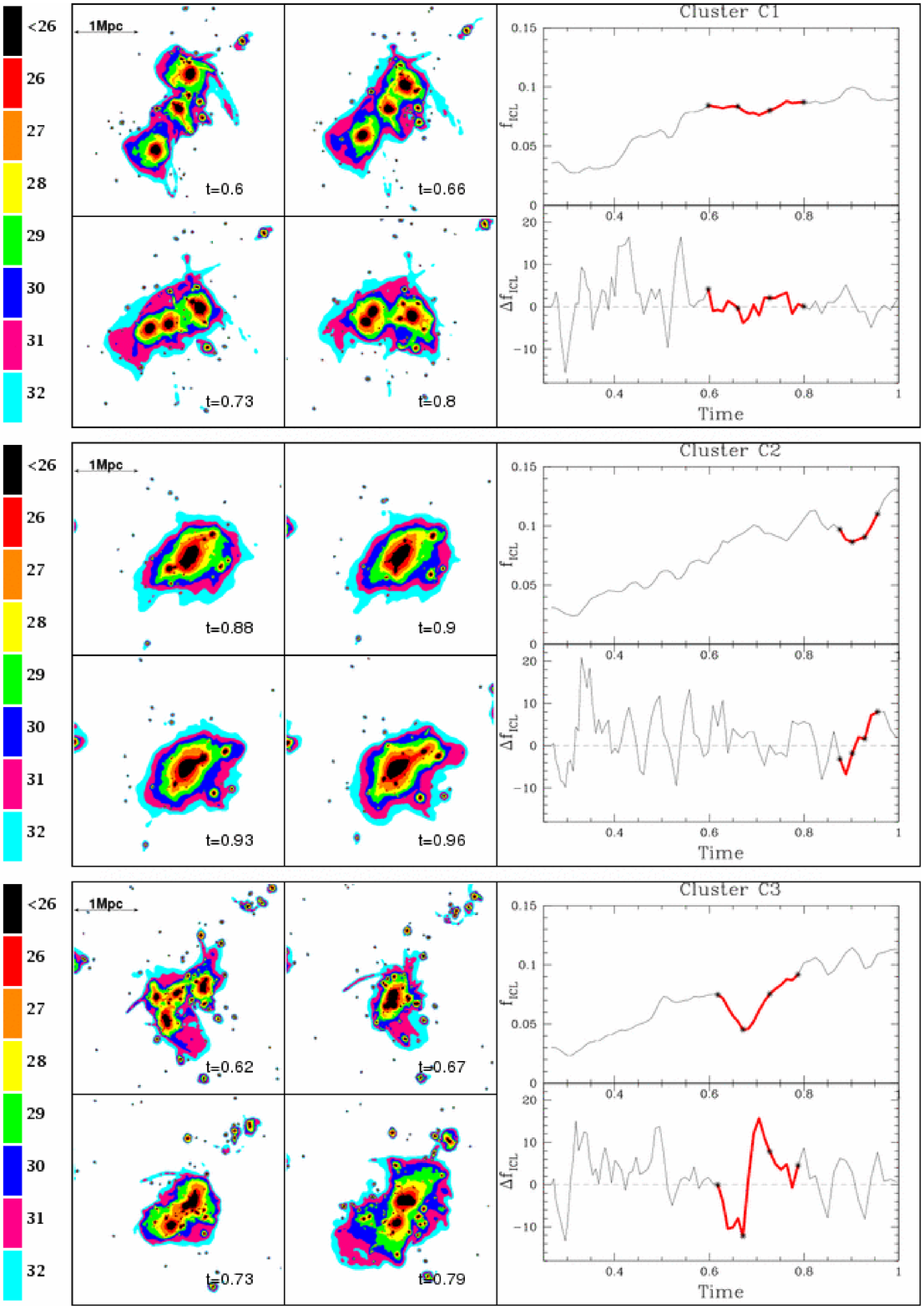}
\caption{
  Simulated images of short segments of the
  evolution of each of our clusters (\textit{top:} C1;
  \textit{middle:} C2; \textit{bottom:} C3) .  The length scale in 
  physical units is shown in the upper left.  The evolutionary time of
  each images is labeled in its lower right corner. The simulated images
  are color-coded by $V$-band surface brightness, as shown by the
  color key at the far left.  Black represents all luminosity at
  $\mu_V < 26.0$ mag/sq.arcsec; each other color  represents a bin of
  one mag/sq.arcsec: red is $\mu_V=26.0-27.0$, orange is
  $\mu_V=27.0-28.0$, etc.  On the right are the $f_{ICL}$ and $\Delta
  f_{ICL}$ plots for each cluster.  The red segments highlight the
  time period shown in the simulated images; the stars mark the
  specific timepoints shown in the images.
   \label{rud:fig:clusters}}
\end{figure}

By following the evolution of our clusters using the simulated images,
shown in Fig. \ref{rud:fig:clusters},
we see that large increases in the ICL luminosity fraction are highly
correlated with small scale interactions and accretion events between
groups within the clusters.  Additionally, $f_{ICL}$ decreases are the
result of temporary increases in the projected density of luminous
material which are a part of the accretion process.  The following
sections show a representative example of this behavior from each
cluster.

\paragraph{Cluster C1}
The top row of Fig. \ref{rud:fig:clusters} shows simulated images of a
short segment of the evolution of cluster C1, along with the $f_{ICL}$
and $\Delta 
f_{ICL}$ functions during this time period.  At $t=0.6$, the cluster
consists of three distinct galaxy groups.  However, at $t=0.8$ these
same three groups still exist within the cluster, relatively
unaltered.  The groups simply orbit one another, resulting in little
accretion or strong inter-group interactions.  During this period, the
fraction of the cluster's luminosity at ICL surface brightness
remains roughly constant.

\paragraph{Cluster C2}
The middle of Fig. \ref{rud:fig:clusters} shows a segment of the
evolution of cluster C2.
Here, we see a significant increase in the ICL luminosity fraction
over a very short time period.  The small group which is below and to the
left of the cluster core at $t=0.88$ plunges through the central mass
concentration, emerging to the upper right.  As the group
enters the core, the projected luminous density increases
temporarily, thus creating the drop in ICL luminosity.  However, as
the group exits the cluster core, the tidal field strips
material from the group, thus moving it to lower surface brightness,
increasing the ICL luminosity fraction

\paragraph{Cluster C3}
In cluster C3, we see our most dramatic example of an ICL production
event, shown at the bottom Fig. \ref{rud:fig:clusters}.  At $t=0.62$ the cluster
consists of no less than four distinct galaxy groups.  At $t=0.67$ all
of these groups have collapsed together, resulting in a very large
loss in the ICL luminosity fraction.  As the groups separate again,
huge amounts of material are stripped from the groups, resulting in
the single largest ICL production event observed in any of these three
clusters.

\section{Conclusions}
By modeling the dynamical evolution of luminous cluster galaxies, we
have been able to trace the evolution of the clusters' diffuse ICL
using simulated broadband images of clusters.  We find that not only
does the fraction of cluster luminosity at ICL surface brightness
increase with dynamical time, but that the ICL evolution is tightly
linked to the specific dynamical histories of the clusters.  Major
increases in ICL luminosity are highly correlated with small scale
interactions and accretions between groups within the clusters.
Because the features of ICL production observed in these simulations
are so tightly correlated with specific evolutionary events within the
clusters, observations of this low surface brightness material has the
potential to reveal a great deal of hitherto inaccessible information
about the dynamical history of galaxy clusters.

This research has been supported by the National Science Foundation,
Research Corporation, and the Jason J. Nassau Graduate Fellowship Fund.

%
%
% BibTeX users please use
% \bibliographystyle{}
% \bibliography{}
%
% Non-BibTeX users please follow the syntax
% the syntax of "referenc.tex" for your own citations
%\input{rudick_references}
%%%%%%%%%%%%%%%%%%%%%%%%%%%%%%%%%%%%%%%%%%%%%%%%%%%%%%%%%%%%%%%%%%%%%%  }

%%%%%%%%%%%%%%%%%%%%%%%%%%%%%%%%%%%%%%%%%%%%%%%%%%%%%%%%%%%%%%%%%%%%%%

\printindex
\end{document}